\documentclass[preprint]{aastex}

\usepackage{amsmath,amssymb,amsfonts}
\usepackage{mathtools}
\usepackage{gensymb}
\usepackage{color}

\shorttitle{Magnetic nulls in space plamas}
\shortauthors{Olshevsky et al.}

\begin{document}

\title{Magnetic null points in kinetic simulations of space plasmas}

\author{Vyacheslav Olshevsky\altaffilmark{1}}
\affil{Centre for mathematical Plasma Astrophysics (CmPA), KU Leuven, Belgium}
\email{sya@mao.kiev.ua}

\and

\author{Jan Deca\altaffilmark{2}}
\affil{Laboratory for Atmospheric and Space Physics (LASP), University of Colorado Boulder} 

\and

\author{Andrey Divin}
\affil{St. Petersburg State University, St. Petersburg, Russia}

\and

\author{Ivy Bo Peng}
\affil{High Performance Computing and Visualization (HPCViz), KTH Royal Institute of Technology, Stockholm, Sweden}

\and

\author{Stefano Markidis}
\affil{High Performance Computing and Visualization (HPCViz), KTH Royal Institute of Technology, Stockholm, Sweden}

\and

\author{Maria Elena Innocenti}
\affil{Centre for mathematical Plasma Astrophysics (CmPA), KU Leuven, Belgium}

\and

\author{Emanuele Cazzola}
\affil{Centre for mathematical Plasma Astrophysics (CmPA), KU Leuven, Belgium}

\and

\author{Giovanni Lapenta}
\affil{Centre for mathematical Plasma Astrophysics (CmPA), KU Leuven, Belgium}

\altaffiltext{1}{Main Astronomical Observatory of NAS, Kyiv, Ukraine}
\altaffiltext{2}{Institute for Modeling Plasma, Atmospheres and Cosmic Dust, NASA/SSERVI, Boulder, Colorado} 

\begin{abstract}
We present a systematic attempt to study magnetic null points and the associated magnetic energy conversion in kinetic Particle-in-Cell simulations of various plasma configurations.
We address three-dimensional simulations performed with the semi-implicit kinetic electromagnetic code iPic3D in different setups: variations of a Harris current sheet, dipolar and quadrupolar magnetospheres interacting with the solar wind; and a relaxing turbulent configuration with multiple null points.
Spiral nulls are more likely created in space plasmas: in all our simulations except lunar magnetic anomaly and quadrupolar mini-magnetosphere the number of spiral nulls prevails over the number of radial nulls by a factor of 3--9.
We show that often magnetic nulls do not indicate the regions of intensive energy dissipation.
Energy dissipation events caused by topological bifurcations at radial nulls are rather rare and short-lived.
The so-called X-lines formed by the radial nulls in the Harris current sheet and lunar magnetic anomaly simulations are rather stable and do not exhibit any energy dissipation.
Energy dissipation is more powerful in the vicinity of spiral nulls enclosed by magnetic flux ropes with strong currents at their axes (their cross-sections resemble 2D magnetic islands).
These null lines reminiscent of Z-pinches efficiently dissipate magnetic energy due to secondary instabilities such as the two-stream or kinking instability, accompanied by changes in magnetic topology.
Current enhancements accompanied by spiral nulls may signal magnetic energy conversion sites in the observational data.
\end{abstract}


\section{Introduction}
\label{sec:intro}

\subsection{Background}
Magnetic nulls, the points in space where the magnetic field vanishes, are believed to be the proxies for magnetic reconnection and magnetic energy dissipation in the Sun \citep{Sweet:1958NC,Longcope:2005LRSP}, and in the Earth's magnetosphere \citep{Xiao:etal:2006NatPh,Wendel:Adrian:2013JGRA,Deng:etal:2009JGRA}.
The interest to the magnetic nulls is growing in the community.
Most recently \citet{Fu:etal:2015JGR} suggested a new method of null point identification based on the first-order Taylor expansion of magnetic field (FOTE).
In distinction from the widely adopted methods \citep{Greene:1992,Haynes:Parnell:2007PhPl}, FOTE could locate nulls even outside the region of space enclosed by the spacecraft.
Using this method, \citet{Eriksson:etal:2015} performed a statistical study of magnetic nulls in the nightside magnetosphere, and concluded that the number of spiral nulls prevailed over the number of radial ones.
The global potential field extrapolations allowed to locate large number of magnetic nulls in the solar corona \citep{Edwards:Parnell:2015SoPh,Freed:etal:2015SoPh}.

The detection of magnetic nulls is only possible when the distribution of the vector magnetic field in space is known.
Vector magnetic field measurements require spectropolarimetric observations or multi-spacecraft measurements (such as those made by Cluster or MMS).
The topology in the vicinity of any stable neutral point can be approximated by the linearization of the magnetic field, which is the base for the null classification.
The eigenvalues of the magnetic field gradient $\mathbf{\nabla}\mathbf{B}$ define whether a null is {\it radial} (degenerates into an X point in 2D) or {\it spiral} (degenerates into an O point) \citep{Lau:Finn:1990,Parnell:etal:1996PhPl}.

In the magnetohydrodynamic (MHD) approximation the theory of magnetic reconnection at null points has been derived by \citet{Pontin:etal:2004,Pontin:Galsgaard:2007JGRA}, and reconnection regimes were classified and generalized by \citet{Priest:Pontin:2009PhPl,Pontin:etal:2011}
Isolated magnetic nulls have been studied by means of MHD \citep{Galsgaard:Pontin:2011b} and kinetic Particle-in-cell (PIC) \citep{Baumann:Nordlund:2012ApJ} simulations.
Observations, however, have suggested that nulls often concentrate in clusters \citep{Deng:etal:2009JGRA,Wendel:Adrian:2013JGRA}.
Several 3D MHD simulations addressed multiple null points \citep{Galsgaard:Nordlund:1997JGR,Wyper:Pontin:2014b,Wyper:Pontin:2014a}.
Notably, the last two publications hinted on the dominating role of the spiral nulls in the models.
\citet{Baumann:Nordlund:2012ApJ} found strong electron acceleration associated with a current sheet and several dynamically evolving nulls in a PIC simulation of solar corona.
\citet{Cai:etal:2006} located and visualized multiple nulls in the PIC simulations of magnetotail reconnection.
They have found a cluster of four nulls, three of which were of spiral topological type.

Previously, we have designed a dedicated three-dimensional magnetic configuration specifically to study nulls in plasma, which exhibited a rather efficient energy dissipation \citep{Olshevsky:etal:2013PhRvL}.
Further investigation has indicated that spiral magnetic nulls and current filaments exhibited strong energy dissipation and magnetic reconnection signatures.
In contrary, energy dissipation events at the radial nulls were rather short-living and localized  \citep{Olshevsky:etal:2015JPP,Olshevsky:etal:2015ApJ}.
These findings motivated us to perform this survey of magnetic nulls in other kinetic simulations of space plasmas.
Our investigation focuses on the relation between spiral nulls, magnetic flux ropes, and regions of intense energy conversion.

\subsection{Simulations}
\label{sec:code}
We perform systematic detection of magnetic null points in different simulations of collisionless plasma carried out with the fully electromagnetic PIC code iPic3D \citep{markidis:etal:2010}.
The code solves the equations of motion for computational particles derived from Vlasov equation, while the time-dependent Maxwell equations for the fields are solved on a stationary grid.
Each computational particle represents a blob of real particles (ions and electrons) that are close to each other in 6D phase space.
The physical units in the code are normalized to the corresponding plasma parameters: proton inertial length $d_i$, proton plasma frequency $\omega_{pi}$, and proton mass $m_i$, hence the magnetic field unit is $m_i\omega_{pi}/e$.

The iPic3D code has been applied to a variety of models: ``classical'' magnetic reconnection \citep{Lapenta:etal:2010PhPl,Divin:etal:2010PhPl}, three-dimensional magnetic reconnection \citep{Vapirev:etal:2013,Lapenta:etal:2015NatPh}, magnetic reconnection at null points \citep{Olshevsky:etal:2015ApJ}, lunar magnetic anomalies \citep{Deca_etal_2015}, planetary magnetospheres \citep{Peng:etal:2015JPP}.
Recently introduced two-way coupling of the iPic3D with the BATS-R-US fluid model \citep{Daldorff:etal:2014} opens even broader perspectives for, e.g., full-scale models of planetary magnetospheres.

We have selected the iPic3D simulations representative of different space plasmas for analyzing magnetic nulls and associated magnetic reconnection.
The three-dimensional paradigm of magnetic reconnection is still in its infancy, thus it is natural to begin with the conventional Harris current sheet simulations (Section~\ref{sec:lapenta}).
More complex initial configuration with asymmetric density profile is studied in Section~\ref{sec:cazzola}.
Magnetic reconnection in fully three-dimensional turbulent configuration with multiple nulls is addressed in Section~\ref{sec:olshevsky}.
Further analysis is devoted to the models where the full set of phenomena arising during the solar wind-magnetosphere interaction happens.
They include: a lunar magnetic anomaly (LMA, Section~\ref{sec:deca}), a dipolar (Section~\ref{sec:bopeng}) and quadrupolar (Section~\ref{sec:divin}) planetary mini-magnetospheres.

Most of the simulations presented in this paper have already been described in other publications.
Not to overwhelm the reader, each section (devoted to the specific initial configuration) describes the setup only briefly, and presents the relevant results in the second part.
The details of all simulations are summarized in the Tables~\ref{tab:allruns} and \ref{tab:magnetosphere}.

\subsection{Location and classification of nulls}
\label{app:poincare}
Locating a magnetic null is essentially the problem of finding a root of a continuous divergence-free vector field.
We use the method of topological degree or Poincar\'e index introduced by \citet{Greene:1992}, with certain improvements as described below.
The topological degree equals to the sum of the amounts of positive and negative nulls inside a closed volume of space.
Its drawbacks are the inability to detect nulls outside this closed volume of space (e.g., a tetrahedron formed by four spacecraft), and impossibility to recover the exact number and positions of the nulls enclosed \citep{Fu:etal:2015JGR}.
However, the grid resolution in kinetic simulations is so fine that it is natural to assume that exactly one null is located in the center of a grid cell with non-zero topological degree.
The advantages of the method are its straightforward implementation and its capability to always detect a null if it is present inside a cell \citep{Haynes:Parnell:2007PhPl}.

To compute a topological degree of the magnetic field enclosed by a tetrahedron (four-spacecraft mission or a `corner' of a hexahedral cell), magnetic field $\mathbf{B}$ is evaluated in its four vertices.
For each triangular facet a solid angle between the three corresponding magnetic field vectors is computed.
The $\mathbf{B}$ vectors are ordered in the right-handed way around the outward normal, therefore the solid angle is given the same sign as their triple product.
This solid angle is essentially the surface of a triangle projected onto a unit sphere in magnetic-field space.
The sum of the four solid angles (the areas of the corresponding spherical triangles) divided by the surface of the unit sphere ($4\pi$) gives the desired topological degree.

The original method \citep{Greene:1992} computes the topological degree over hexahedral boxes divided into 12 triangles.
We prefer to use tetrahedra instead, to keep our implementation consistent with the spacecraft observations \citep{Xiao:etal:2006NatPh,Deng:etal:2009JGRA}.
Each cubic computational grid cell is subdivided into four non-overlapping tetrahedrons with the vertices at the grid nodes.
The topological degree is evaluated for each tetrahedron, and, when a null is encountered, its topological type is detected.
The classification is based on the eigenvalues $\lambda_1,\,\lambda_2,\,\lambda_3$ of the gradient of the magnetic field $\mathbf{\nabla}\mathbf{B}$ in the vicinity of the null.
We estimate $\mathbf{\nabla}\mathbf{B}$ using the first-order Taylor expansion of the field around the tetrahedron's center of mass as described by \citet{Khurana:etal:1996IEEE}.
The null is positioned at the center of mass of the corresponding tetrahedron.

An advantage of our implementation is the formula for the solid angle proposed by \citet{VanOosterom:Strackee:1983IEEE}
\begin{equation}
\tan\left(\frac{\Omega}{2}\right)=\frac{\mathbf{R}_1 \cdot \left[ \mathbf{R}_2 \times \mathbf{R}_3 \right]}{R_1 R_2 R_3 + \left(\mathbf{R}_1\cdot \mathbf{R}_2\right)R_3 + \left(\mathbf{R}_1\cdot \mathbf{R}_3\right)R_2 + \left(\mathbf{R}_2\cdot \mathbf{R}_3\right)R_1},
\end{equation}
where $\mathbf{R}_1$, $\mathbf{R}_2$ and $\mathbf{R}_3$ are the vectors enclosing the solid angle.
This formula is faster and more convenient than the traditional implementation based on the Cosine theorem. 
In particular, there is no need for zero-denominator checks in modern programming environments: all errors are handled by the arctan2 function.

Our null classification follows the one introduced by \citep{Lau:Finn:1990}, based on the eigenvalues of the $\mathbf{\nabla}\mathbf{B}$.
Due to the divergence-free nature of the magnetic field, the condition $\lambda_1+\lambda_2+\lambda_3=0$ should be satisfied.
Thus the following cases are possible :
\begin{enumerate}
\item One eigenvalue turns to zero (in reality the condition is $|\lambda_1|<\sigma$, where $\sigma$ is a small positive number). Such null is essentially two-dimensional, and is unstable in 3D.
Two types of 2D null points are possible: (1) X-point, when $\lambda_2,\,\lambda_3$ are real; (2) O-point, when $\lambda_2,\,\lambda_3$ are complex conjugates.
\item All three eigenvalues are non-zero and real, the null is of {\it radial} type: either A (negative, two eigenvalues are negative) or B (positive, two egenvalues are positive). X-point is a degenerate case of the radial null.
\item One eigenvalue is real, and two others are complex conjugates. This combination defines a {\it spiral} null: either As (negative) or Bs (positive). O-point is a degenerate case of the spiral null.
\end{enumerate}

Theory predicts conservation of the topological degree, i.e., each newly created or just disappeared positive null must have its negative counterpart, and vice versa.
However, this is not always the case in our simulations: at certain moments the number of positive nulls deviates from the number of negative nulls.
In practice errors may occur on the stage of null detection due to the noise in magnetic field measurements and the finite resolution of the numerical grid or spacecraft instruments.
Additional errors are introduced during null classification, especially when the eigenvalues of the $\mathbf{\nabla B}$ have very small real or imaginary parts.
The influence of these errors on the observations of null points are addressed in \citet{Eriksson:etal:2015}.
In particular, it was found that a null point is more likely to flip its ``sign'', e.g., A to B, or As to Bs.
Change of the spiral or radial ``nature'' of the null is less probable.

The summary of all nulls found in our simulations is presented in Table~\ref{tab:allnulls}.

\section{Three-dimensional Harris current sheet}
\label{sec:lapenta}
The Harris equilibrium (aka Harris current sheet) \citep{Harris:1962NC} is not  directly representative of any specific physical system.
But it has proven abilities to describe as a first approximation a large array of systems with oppositely directed magnetic fields.
Applications include magentospheric, coronal, laboratory and astrophysical plasmas (see \citet{Yamada:etal:2010RvMP,Treumann:Baumjohann:2013}, and references therein).
The vast majority of models of magnetic reconnection studied within last 50 years consider Harris equilibrium as an initial condition.
In our setup the initial magnetic field $B_x(y)=B_0 \textrm{tanh}(y/L)$ is directed along the X axis, the current is along Z: $j_z=j_0 \textrm{cosh}^{-2}(y/L)$. 

An initial perturbation is added to vector potential to create a designed X-line \citep{Lapenta:etal:2010PhPl}
\begin{equation}
\delta A_z = A_{z0} \cos\left(2\pi x/L_\Delta\right) \cos\left( \pi y / L_\Delta \right) e^{-\left(x^2+y^2\right)/\sigma^2},
\label{eq:perturbation}
\end{equation}
where $L_\Delta=10\sigma$ and $\sigma=d_i/2$.
The initial state, including the perturbation, is initially invariant along Z where periodic boundary conditions are used.
In the X and Y coordinates the plasmas and fields can freely cross the boundary via open boundary conditions \citep{Wan:Lapenta:2008}.

While in some cases, as the Earth magnetotail, an exact field reversal is an interesting case, in most other situations reconnection happens via component reconnection \citep{Fuselier:etal:2011JGRA}.
In this case the field does not vanish to zero in the central current layer but rather a guide field is present in the Z direction: this field is referred to as guide field.
For this reason, a {\it guide field} is added to the initial Harris equilibrium directed along the current.
We consider two cases here: a simulation without a guide field, and a simulation with a small guide field $B_g=0.1B_0$.
Higher guide fields tend to suppress nulls and are not considered in the present study.

The simulations are carried out with an initial Harris equilibrium with a uniform plasma background of $n_b=0.1n_0$ of the peak Harris density. 
Further details of the two simulations are given in the Table~\ref{tab:allruns}.

\subsection{Results: Harris}
\begin{figure}
\centering
\includegraphics[width=1\columnwidth]{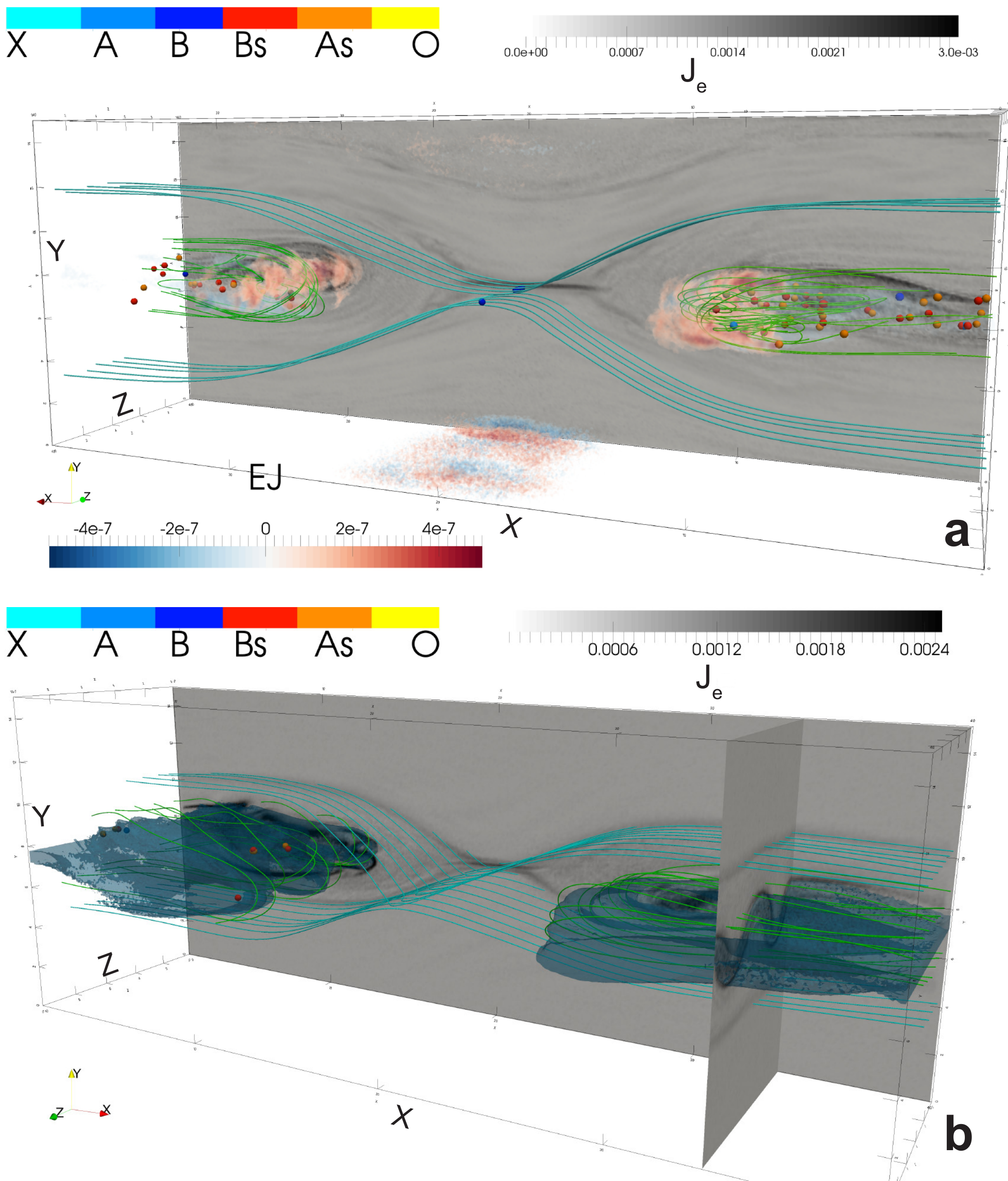}
\caption{Magnetic nulls in a three-dimensional Harris sheet configuration: (a) a simulation with zero guide field, (b) a simulation with $B_g=0.1$.
In this and all other figures nulls are depicted by spheres, their color-coding is the same throughout the paper.
Cyan and green field lines depict magnetic topology, black and white slices in the background indicate electron current density $j_e$.
In (a) the energy conversion $\vec{E}\cdot\vec{j}$ is indicated with the red and blue volume rendering (fog-like). 
Some energy dissipation adjacent to the bottom and upper boundaries is an artifact.
In (b) blue semi-transparent isosurfaces indicate regions of high electron density (pileup region, dipolarization front).
}
\label{fig:lapenta1}
\end{figure}

For our analysis we have selected one snapshot from each simulation taken at $t=2000\omega_{pi}^{-1}$, when a number of magnetic nulls emerge in the guide field run.
At this stage, the initial magnetic topology has undergone significant change, and well defined reconnection exhausts have formed. 
High density pileup regions have been created on the left and on the right from the central X-line, already non-planar, perturbed by a warping visible in semi-transparent electron density isocontours in Fig.~\ref{fig:lapenta1}b.
These pileup regions are also characterized by major energy conversion (blue and red shade in Fig.~\ref{fig:lapenta1}a).
The energy dissipation and density distribution is qualitatively very similar in the two cases, and for clarity each feature is emphasized only on one image.

In both cases nulls concentrate in the pileup regions.
The number of nulls in the simulation without guide field is notable larger, and a few radial nulls are found on the ``X-line".
Although these nulls are not associated with energy conversion, they are proxies for the global changes of the magnetic topology (by reconnection), hence large-scale energy release (Fig.~\ref{fig:lapenta1}a).
In the second snapshot nulls only appear on the left side (Fig.~\ref{fig:lapenta1}b). 
This asymmetry, although not important for the present analysis, is, ruined at a later stage in the simulation.
The overall number of radial nulls is much smaller than the number of spiral nulls (Table~\ref{tab:allnulls}).

\begin{figure}
\centering
\includegraphics[width=1\columnwidth]{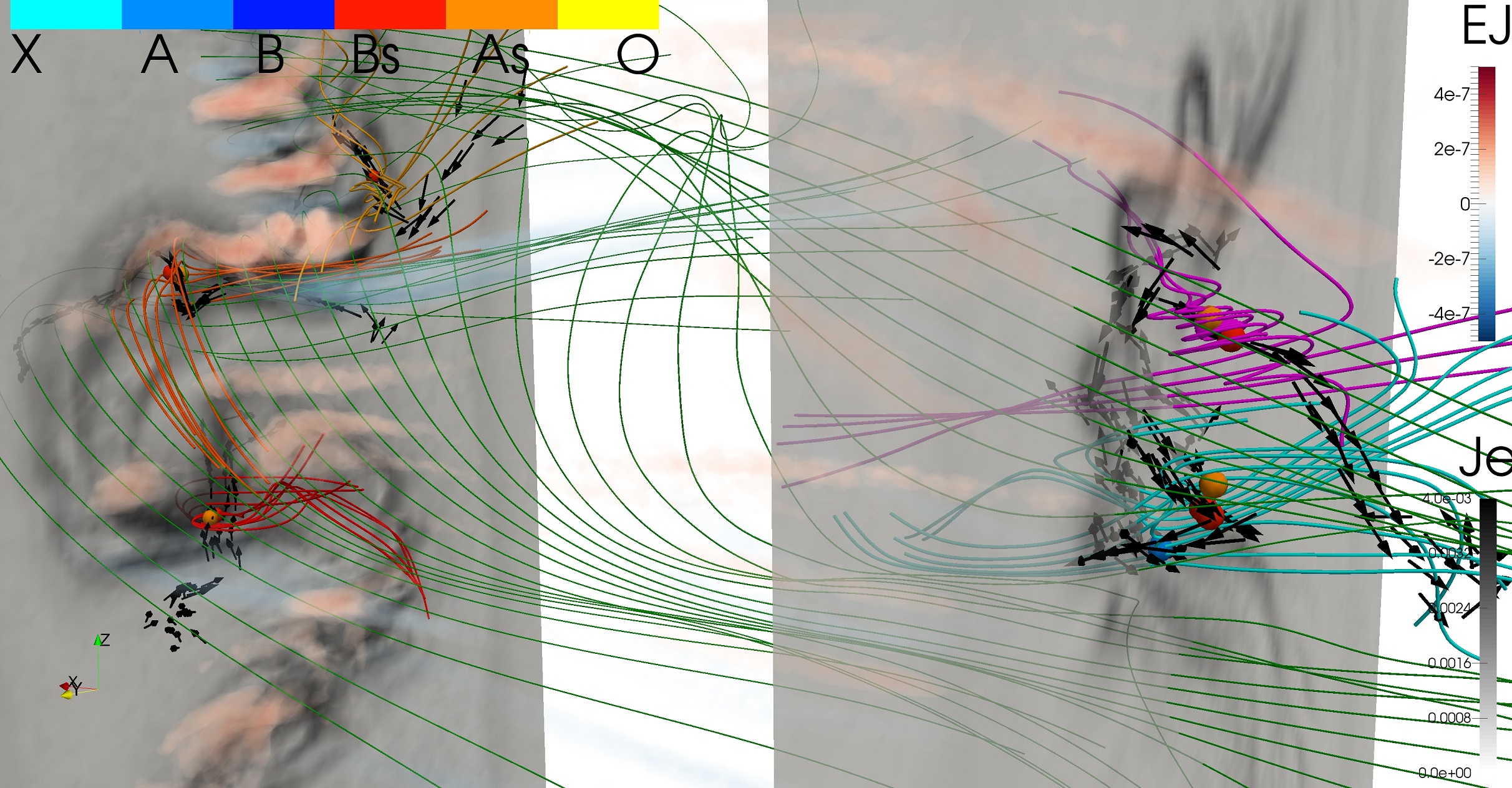}
\caption{Zoom on the nulls in the Harris current sheet simulation with $B_g=0.1$. 
Nulls aren't evenly distributed, they lie close to the two grayscale planes perpendicular to the X axis; the color scale denotes electron current density $j_e$.
Electron current streamlines bypassing the selected nulls are shown with black arrows.
The energy conversion $\vec{E}\cdot\vec{j}$ is indicated with the red and blue volume rendering (fog-like). 
Green, cyan, orange, magenta and red field lines denote magnetic topology in the vicinity of nulls.
}
\label{fig:lapenta2}
\end{figure}

Major energy dissipation in both simulations happens in the pileup regions in the reconnection exhaust marked by darker color in the background slices in Figure~\ref{fig:lapenta1}. 
There are clusters of nulls that are colocated with energy conversion regions, and there are those that do not exhibit dissipation.
A zoom on the simulation domain enclosing the nulls in the simulation with $B_g=0.1$ is shown in Figure~\ref{fig:lapenta2}.
Interestingly, all nulls are concentrated close to the two planes perpendicular to the X axis: they are colored with grey-black colormap showing electron current density $j_e$.
Warping and kinking of the current sheet produces twisting of the ambient magnetic field lines (green), and results in the formation of the null points.
This mechanism resembles topological bifurcations \citep{Murphy:etal:2015} observed in the MHD simulation of a null point current sheet by \citet{Wyper:Pontin:2014a}.
Earlier \citet{Lapenta:etal:2015NatPh} have found that such ``secondary reconnection sites", formed in the pileup regions with twisted magnetic fields, might be important for energy conversion.

Observing more closely  the magnetic nulls, it emerges how they are enclosed in the twisted magnetic field lines (pink, red and yellow in Fig.~\ref{fig:lapenta2}), resembling  vortices produced by turbulence.
Electron current streamlines through the nulls are shown by the black vectors. 
Nulls form close to where these streamlines bend.
There is no direct evidence for the causal relationship between the nulls and energy dissipation.
It seems, however, that nulls in the reconnection pileup region are created when magnetic topology changes due to non-linear interaction of the current filaments.
These filaments may form due to various instabilities in the initial current sheet.
Current sheet breakup mechanism is a classical concept, the breakup due to tearing is described, by, e.g., \citet{Galeev:Sudan:1984}.

The topology of the magnetic nulls in the simulation without guide field is more complicated because the number of nulls is higher.
However, qualitatively the situation is very similar, and allows us to conclude: (1) nulls are created in the reconnection pileup regions as secondary topological bifurcations due to interacting current filaments; (2) most of the nulls are of spiral type; (3) radial nulls at the initial ``X-line" do not excibit energy dissipation.

\section{Asymmetric reconnection in double current sheet}
\label{sec:cazzola}
This section analyses the null evolution in a 3D simulation of asymmetric reconnection.
The asymmetric configuration reproduces the situation observable at the dayside magnetopause, where the dense and low magnetic field shocked solar wind passing through the bow shock encounters the low density high field magnetospheric plasma, eventually leading to a reconnection event when the solar wind magnetic field is southward \citep{phan1996,mozer2002,mozer2008,mozer2008themis}. 
The initialization setup introduced by \citet{cazzola15} for a 2.5D study is extended here to the third dimension. 

The upper layer considers the continuous hyperbolic profiles typically adopted in literature \citep{pritchett2008}, while the lower layer describes a pure tangential discontinuity set with an extremely steep gradient. 
Therefore, the total profile reads
\begin{equation} \label{eq:Beq}
 B_x\left(y\right) =
   \begin{dcases}
     \frac{B_0}2{}, \quad \text{$y \le \frac{L_y}{4}$} \\
     B_0 \left[ \tanh \left(\frac{y - y_2}{\lambda} \right) + R \right],   \quad \text{$y > \frac{L_y}{4}$}
   \end{dcases}
\end{equation}

\begin{equation} \label{eq:neq}
  n(y) = 
   \begin{dcases}
    n_0, \quad \text{$y \le \frac{L_y}{4}$} \\
    n_0 \left[ 1  - \alpha \tanh \left(\frac{y - y_2}{\lambda} \right) \right] \\
    - \alpha \tanh^2 \left( \frac{y - y_2}{\lambda} \right),  \quad \text{$y > \frac{L_y}{4}$}
   \end{dcases}
\end{equation}
where $y_2 = 3L_y/4$, and the current sheet thickness is $\lambda=0.5$. 
To satisfy equilibrium between the two different plasmas across the current sheet, $R = 0.5$ and 
$\alpha = 0.33$. No guide field is set in this simulation.


Having two inversion layers (at $y=L_y/4$ and $y=3 L_y/4$) in a single simulation offers the possibility to readily compare the two evolutions under different initial conditions.
In both inversion layers presented here the asymptotic magnetic field in the $x$ direction jumps from $B_x=-B_0/2$ to $B_x=3/2 B_0$. 
Conversely, the density increases from $n=n_0/3$ to $n=n_0$.
The denser, low magnetic field area is located at $L_y/4 < y < 3 L_y/ 4$. 
Boundary conditions are periodic for fields and particles in all directions.
The lower inversion layer transitions abruptly between the dense and the thin areas: a strong discontinuity (i.e., a step function) is used to model the magnetic field and density profile across $y= L_y/ 4$ \citep{ferraro1952theory,biernat1989structure}. 
No current profile is initially set up, but a strong current layer is soon generated self-consistently by particle motion. 
The upper inversion layer is initialized with the ``gentler'' magnetic field and density profile \citep{pritchett2008}) given by Equations (1) and (2) in \citet{cazzola15}. 
A consistent out of plane current is initialised assuming that all current is carried by electrons.   

A perturbation with the same profile as in Section~\ref{sec:lapenta} (Equation~\ref{eq:perturbation}) is initially applied to the upper layer, while the lower layer is not perturbed: its intrinsically unstable configuration causes reconnection to start with no external intervention. 
The upper layer can be used to contrast the Harris sheet simulation results of the previous section. 
The lower layer can be used to study null evolution during island coalescence at the magnetopause.

\subsection{Results: asymmetric}

\begin{figure}
\centering
\includegraphics[width=1\columnwidth]{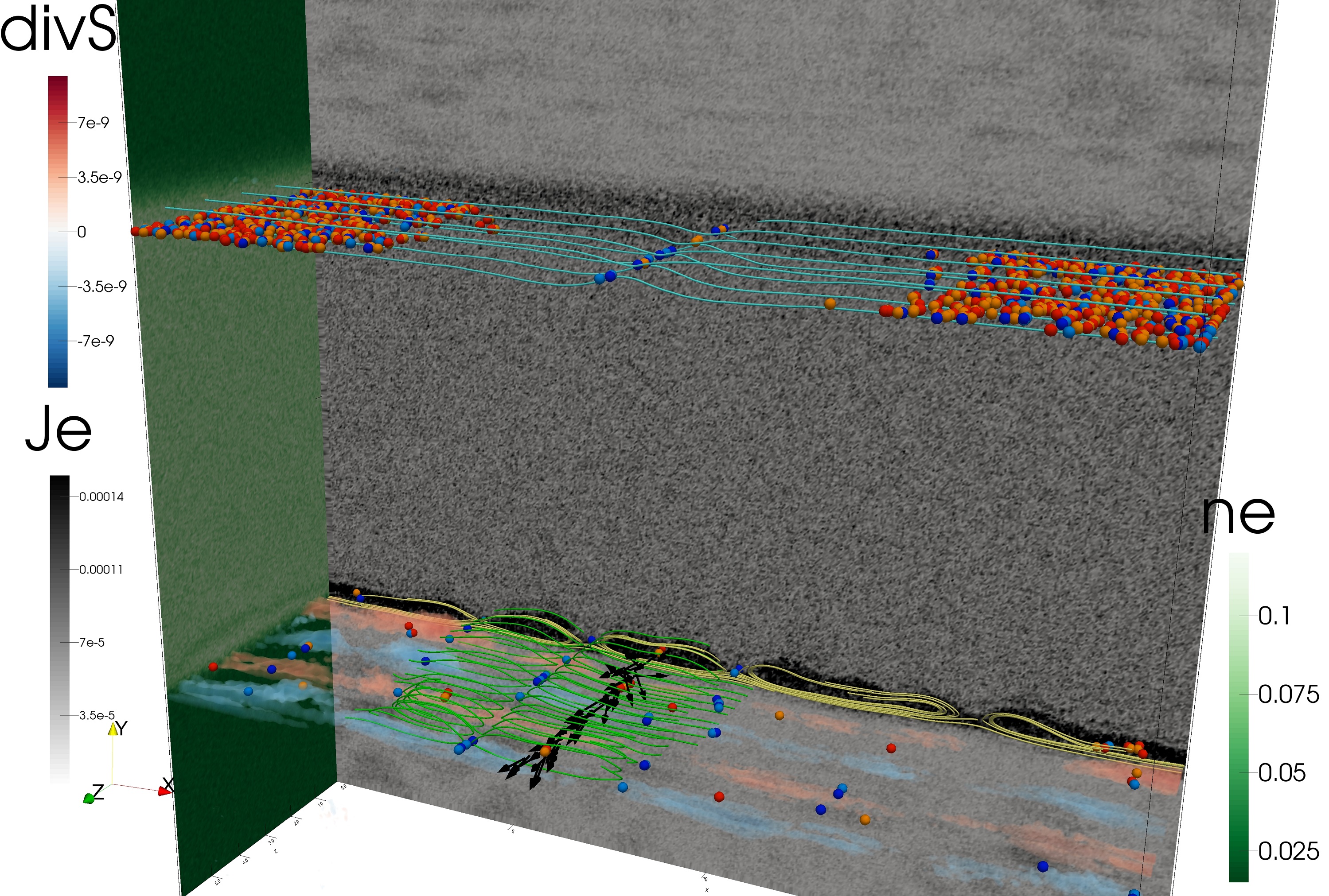}
\caption{
Overview of the two inversion layers evolution in the asymmetric reconnection simulation. 
The white-green YZ slice denotes asymmetric electron density distribution.
The white-black XY slice denotes electron current density: 5 magnetic flux ropes have formed in the bottom inversion layer, also highlighted by yellow magnetic field lines, resembling magnetic islands in the XY plane. 
Black arrows show electron current vectors along a spiral null line inside one such island.
Nulls are shown with color spheres with the conventional color coding (Fig.~\ref{fig:lapenta1}).
Blue and red fog-like volume rendering shows another proxy for energy conversion, the divergence of Poynting vector $\mathbf{\nabla}\cdot\mathbf{S}$.
}
\label{fig:asym}
\end{figure}

In Figure~\ref{fig:asym} an overview of the simulation at time $\omega_{pi}t=4500$ is given.
The two inversion layers are noticeable in the transitions in electron density (green slice), current density (black-white slice), and nulls concentrations.
In the upper layer, a single X line is present, as per initial perturbation.
Interestingly, several spiral nulls are detected in the X line, either real topological bifurcations or topological type misdetections (see Appendix~\ref{app:poincare}).
In the not yet perturbed regions of the upper layer dozens of radial and spiral nulls are found, created by the break-up of the non-generic initial null surface due to numerical noise.
The upper current sheet is rather thick, the current density is low, and no major energy conversion is associated with it at this stage of the evolution.

In the lower layer, consistent with \citet{cazzola15}, five magnetic islands (magnetic flux ropes in 3D) have formed. 
They are highlighted in Figure~\ref{fig:asym} with yellow field lines and enhanced electron current density.
Two typical flux ropes and an X-line separating them are illustrated by the green field lines. 
In the center of each magnetic flux rope there is a current streaming along a null line that consists of spiral nulls (O-line), demonstrated with black arrows showing electron current vectors.
This is a typical picture of tearing of the initial current sheet \citep{Biskamp:2000,Galeev:Sudan:1984}.

Unlike the Harris sheet simulations reported above the asymmetric simulation doesn't have background species, and the current profile is rather noisy (black-and-white slice in Fig.~\ref{fig:asym}).
Therefore as a proxy for energy conversion the value of the Poynting vector divergence $\nabla\cdot\mathbf{S}$ is used in Figure~\ref{fig:asym}.
Energy dissipation is apparent below the flux ropes, in the less dense plasma region.
The most prominent feature of the $\mathbf{\nabla}\cdot\mathbf{S}$ distribution is, however, a characteristic alternation of positive and negative stripes in the Z direction. 
This hints to the presence of a current aligned instability with wavelength $\lambda \sim L_z/2$ (this value is most probably affected by the box size in the Z direction) responsible for the direction of the energy flow and for the alternative energy exchange between fields and particles. 
The natural candidate (see \citet{divin2015lower} and references therein) is the Lower Hybrid Drift Instability, LHDI \citep{lapenta2002nonlinear, daughton2003_LHDI_emBranch}. 
The LHDI has been shown to be able to exchange momentum between ions and electrons at least at the flanks of symmetric current sheets through electrostatic fluctuations \citep{innocenti2007momentum}. 
Observations of LHDI signatures in asymmetric reconnection simulations are reported in literature (\citet{pritchett2011}, and references therein).
In asymmetric reconnection, the characteristic LHDI rippling is observed mainly in the separatrices bordering the lower density regions, i.e. the lower separatrices in the lower inversion layer.

The asymmetric reconnection scenario confirms and extends the conclusions made from the analysis of magnetic nulls in the classical Harris sheet configurations: (1) nulls are created inside magnetic flux ropes, along with current filaments; (2) spiral nulls dominate over radial ones; (3) often nulls are not associated with major energy conversion.

\section{Relaxing configuration with multiple nulls}
\label{sec:olshevsky}
The magnetic field configuration with multiple nulls, proposed by \citet{Olshevsky:etal:2013PhRvL}, reveals a very high magnetic energy dissipation rate.
The setup is fully periodic:
\begin{eqnarray*}
  B_x & = & -B_0\cos{\frac{2\pi x}{L_x}}\sin{\frac{2\pi y}{L_y}}, \qquad\qquad \\
  B_y & = & B_0\cos{\frac{2\pi y}{L_y}}\left( \sin{\frac{2\pi x}{L_x}} - 
        2 \sin{\frac{2\pi z}{L_z}} \right), \\
  B_z & = & 2 B_0\sin{\frac{2\pi y}{L_y}}\cos{\frac{2\pi z}{L_z}}, \qquad\qquad \\
\label{eq:initial}
\end{eqnarray*}
where the simulation domain extent is $L_x=L_y=L_z=20$, the magnetic field amplitude is $B_0=0.0127$, and the particle density is $n_0=1$ in code units (Section~\ref{sec:code}).
The initial particle velocity distribution is a Maxwellian with ion/electron temperature ratio $T_i/T_e=5$.
The simulation passes through the stages with different plasma $\beta$: it begins with the magnetic/thermal energy ratio $W_{mag}/W_{th}=1.38$, and ends with $W_{mag}/W_{th}=0.07$.
Under such conditions the electron inertial and gyration scales are well resolved.
This simulation has been analyzed in greater details by \citet{Olshevsky:etal:2015ApJ}, here we only describe the energy conversion processes relevant to the spiral and radial nulls.

\subsection{Results: ``multiple nulls''}
\begin{figure}
\includegraphics[width=1\textwidth]{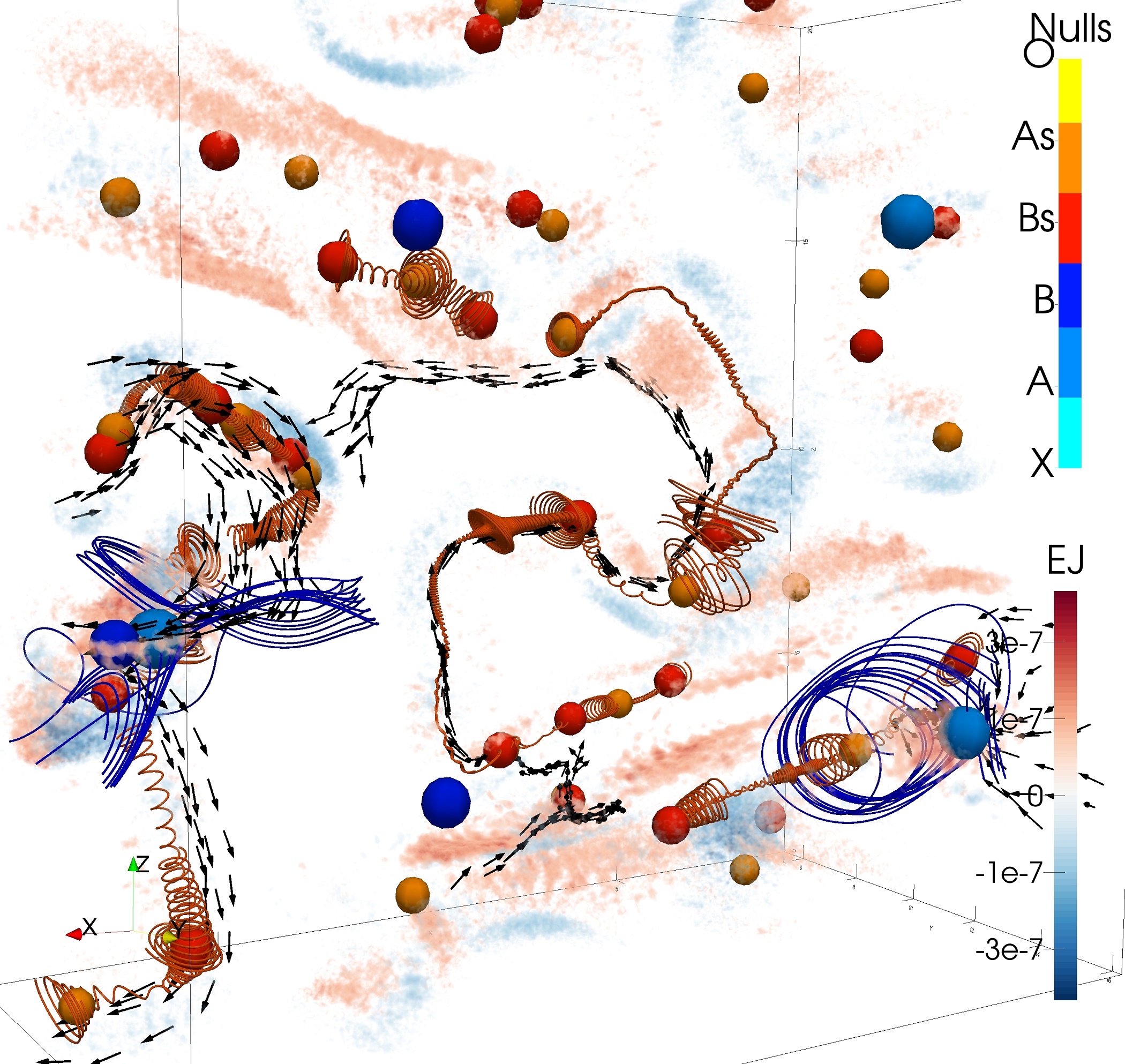}
\caption{A snapshot from the ``multiple nulls'' simulation. 
Conventional indicators of nulls (color spheres) and energy dissipation measure $\mathbf{E}\cdot\mathbf{j}$ (blue-red fog-like volume rendering) are shown.
Most dissipation is attributed to non-linear interaction of the current channels (black arrows show electron current vectors) embedded into twisted magnetic fields.
Orange field lines show magnetic topology along such flux ropes: null lines with alternating As and Bs nulls.
Sometimes, however, energy dissipates in the radial nulls, as happens with a short-living A-B null pair in the left surrounded by the blue field lines.
\label{fig:olshevsky1}}
\end{figure}

The magnetic field configuration described by Equations~\ref{eq:initial} is divergence-free, but not force-free.
Because the initial particle density is uniform, the forces at $t=0$ are not balanced.
Pressure imbalance is a trigger for the initial explosive relaxation of the system.
During this phase half of the magnetic energy is converted to the particle energy \citep{Olshevsky:etal:2015JPP}.
The second stage of the evolution is characterized by randomly-distributed, turbulent dissipation events taking place all over the domain.
The dominating majority of nulls in this simulation are spiral (Figure~\ref{fig:olshevsky1}, Table~\ref{tab:allnulls}).
In part, this is the legacy of the setup: the initial configuration contains nine null lines consisting of O-points.

A slab of the simulation domain that includes different nulls at $\Omega_{ci}t=42.5$, when the initial explosive relaxation was already over, is shown in Figure~\ref{fig:olshevsky1}.
Orange magnetic field lines are twisted around the current channels shown with black electron current density vectors. 
Positive (Bs) and negative (As) spiral nulls alternate along these spiral null lines.
A complex warped topology makes identification of the separators of the adjacent nulls tricky, and it is possible they lie on the intersecting fan surfaces. 
However, \citet{Murphy:etal:2015} suggest that such nulls might not be connected by a separator just after their emergence or just before their disappearance.
Helical field lines of the fans connect the nulls with the external magnetic field.
Such topology is very similar to the one deduced from the Cluster observations of a null pair in the Earth's magnetosheath by \citet{Wendel:Adrian:2013JGRA}. 
Another null pair, observed by \citet{Deng:etal:2009JGRA}, has shown quite large angle between the spines of the two nulls, and a fan-fan separator line.

Interactions of the adjacent flux ropes may lead to spontaneous creation of nulls.
For example, a short-lived pair of radial nulls in the left part of Figure~\ref{fig:olshevsky1}.
The distance between these two nulls is $1\,d_i$, they live for a few ion gyration times, and are accompanied by strong currents and enhanced energy conversion.
A spontaneous emergence of magnetic nulls was explained in terms of topological bifurcations by \citet{Wyper:Pontin:2014b}.
The interconnection between the emerged radial nulls is more complex than between the spiral nulls in a flux rope.
The field lines forming the fans and the spines of the nulls are bent and twisted (blue field lines in Fig.~\ref{fig:olshevsky1}).
The fan of the A null and the spine of the B null are formed by the same group of field lines that start from the left (X) boundary of the simulation domain.
These field lines encircle two adjacent flux ropes (null lines), joining their topologies.
Reconstruction of such topology from observations is not conceivable in the linear approximation.
The current structure is complex as well: the stream of electrons approaches the B null, where it bends towards the A null, and finally is scattered away forming a small-scale current sheet.
Presence of a short-lived null pair, a complex current structure and an enhanced energy dissipation rate $\mathbf{E}\cdot\mathbf{j}$ are the indicators of this magnetic reconnection event.

Despite the aforementioned enhanced energy release at the radial null pair, the $\mathbf{E}\cdot\mathbf{j}$ (blue and red fog-like volume rendering in Fig.~\ref{fig:olshevsky1}) is more prominent around the magnetic flux ropes.
Namely, the dissipation happens where the current channels bend, and where the adjacent currents channels interact with each other.
Two-stream instabilities cause energy conversion on the interfaces of the oppositely directed current channels \citep{Olshevsky:etal:2015JPP}.
Unlike the Harris equilibrium scenarios (Sections~\ref{sec:lapenta}, \ref{sec:cazzola}), the current channels here are not a product of the current sheet filamentation, but rather artifacts of the setup.
Nevertheless, the energy conversion patern, and the dominance of the spiral nulls are in qualitative agreement with the conclusions drawn in the previous sections.
This, rather unexpected, result hints to the common nature of the processes governing energy conversion in various types of turbulent space plasmas.
Main actors of these processes are interacting current filaments or magnetic flux ropes.
Spiral nulls form and disappear along the axes of these flux ropes; no specific energy release events are associated solely with these nulls.
Short-living radial null pairs form between interacting flux ropes, and cause localized energy dissipation.

\section{Lunar Magnetic Anomaly (LMA)}
\label{sec:deca}
In 1959 the Luna 2 mission made the (at the time) surprising discovery that our Moon does not support a global magnetic field~\citep{Dolginov_etal_1960}.
The apollo missions shortly after, however, did detect small areas of weak crustal magnetic fields~\citep{Dyal_etal_1970, Dyal_etal_1974, Russell_etal_1974, Sharp_etal_1973, Fuller_1974}.
Recent high-resolution measurements characterised these lunar magnetic anomalies (LMAs) to have size up to several 100\,km with surface magnetic field strengths ranging from 0.1 up to 1000\,nT~\citep{Lin_etal_1998, Mitchell_etal_2008, Richmond_Hood_2008, Purucker_2008, Purucker_Nicholas_2010}. 

Since LMAs are rather tiny compared to typical ion plasma scales in the solar wind, the solar wind - LMA interaction is dominated by highly non-adiabatic physical processes~(e.g.,~\citealt{Deca_etal_2014,Deca_etal_2015,Howes_etal_2015}, and reference therein).
Even more, in situ Kaguya and Chandrayaan measurements have indicated that some of these crustal fields might be able to locally shield the lunar surface from direct impact by the solar wind plasma and form a so-called 'mini-magnetosphere'~\citep{Lin_etal_1998,Wieser_etal_2010, Saito_etal_2010, Vorburger_etal_2012}.



The ability to investigate the effects of charge separation, hence a full-kinetic model, are by principle a must for detailed modelling of the near-surface lunar plasma environment. \citet{Deca_etal_2014,Deca_etal_2015} observe in their 3-D simulations the formation of a mini-magnetosphere above a dipolar magnetic field configuration resembling the strongest component of the Reiner Gamma anomaly~\citep{Kurata_etal_2005}. 
The configuration is highly driven by electron motion, displays reflection and scattering of ions, and heats and deflects electrons perpendicular to the magnetic field under a influence of an $\mathbf{E}\times\mathbf{B}$-drift mechanism (See also Figure~\ref{fig:LMA} where typical streamlines for both species in the simulation are shown in harmony with the 2-D surface density profile).

The setup used here is identical to the run A in~\citet{Deca_etal_2015}, and is representative of the solar wind -- LMA interaction under quiet solar wind conditions at 1~a.u.. 
The solar wind velocity is directed perpendicular to the lunar surface, whereas the interplanetary magnetic field (IMF) is parallel to the surface and directed along the dipole axis, see also Tables~\ref{tab:allruns} and \ref{tab:magnetosphere} for more details.

\subsection{Results: LMA}
\begin{figure}
\centering
\includegraphics[width=1\columnwidth]{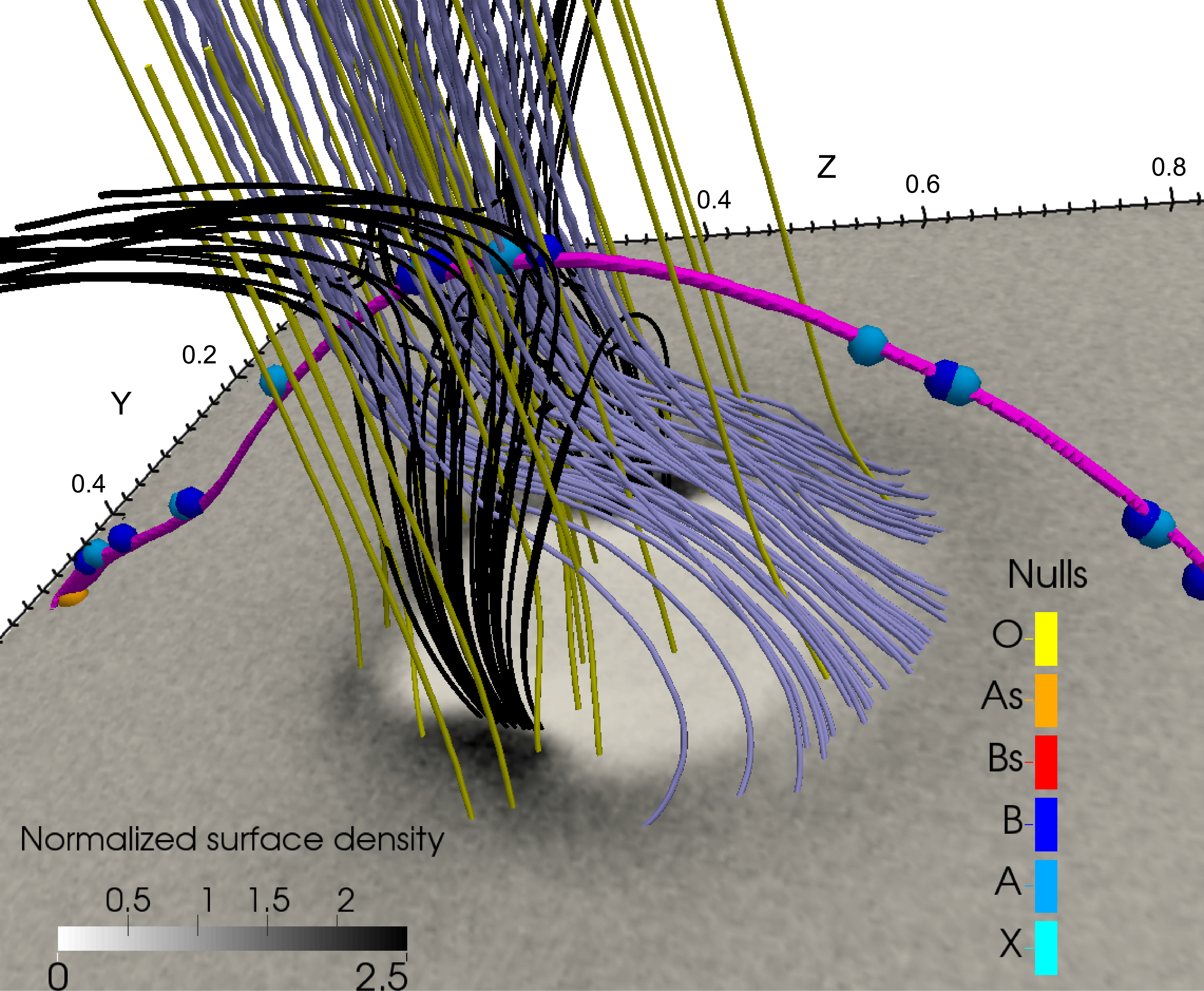}
\caption{Null point identification along the null line overspending the LMA mini-magnetosphere structure. We show the normalised 2-D surface density profile, magnetic field lines (black) and typical electron (purple) and ion (ocher) streamlines. The null points are indicated with coloured spheres. To improve visual conceptualisation of the null line, areas with close-to-zero magnetic field are shown in pink.}
\label{fig:LMA}
\end{figure}

The LMA scale-size is small with respect to the solar wind ion-gyroradius ($r_i \sim 1.5\,d_i > L_{box}$) and consequently no clear shock associated with the mini-magnetosphere structure is observed, in contrast to larger magnetospheres (Sections~\ref{sec:bopeng} and \ref{sec:divin}). 
No stationary shock can exist as plasma penetrates the halo at a too high speed. 
An arc of zero total magnetic field is created in the magnetic configuration across the entire structure at $0.27\,d_i$ above the surface (measuring its highest point) by choosing the IMF direction parallel to the dipole moment. 
This null line consists essentially of radial nulls (Figure~\ref{fig:LMA}).
The dipole centre is located at $0.1\,d_i$ below the absorbing outflow boundary representing the lunar surface under idealised conditions, that is ignoring surface charging and secondary particle effects. 
Small-scale kinetic instabilities are present along both the density halo and the null line and can most probably be attributed to a mirror instability \citep{Deca_etal_2014}. 

Magnetic nulls in this simulation are only found on the aforementioned null line, and are only of radial type: no spiral nulls are found in any of the simulation snapshots.
Commonly used indicators of magnetic reconnection and energy dissipation or particle flows are not observed. 
However, the nulls are unstable over time and emerge/disappear continuously at different locations along the null line due to the symmetry in the initial configuration.
No signatures of magnetic reconnection is observed because of the small, electron scales of the magnetic configuration.
Hence, in this case the null points have a potential nature and are simply topological features rather than the indicators of a fully developed magnetic reconnection process.
Most likely the solar wind speed is too high for typical magnetic reconnection to form under the current topology.

\section{Dipolar mini-magnetosphere}
\label{sec:bopeng}
Global PIC simulations of magnetospheres allow us to study basic plasma physics from first principles in realistic magnetic field topologies and with self-consistent distribution functions. 
For example, our previous works include simulations of magnetic reconnection \citep{Peng:etal:2015JPP} and bow shock formation \citep{peng2015bowshock}. 
In these simulations, solar wind electrons and ions are injected from one side of the simulation box while particles and plasma waves exit from the other sides. 
These global PIC simulations are still limited to a box size of a few hundred ion inertial lengths and non-realistic ion-to-electron mass ratios because of the high computational cost.

The solar wind interaction with the dipolar magnetic field of a planet forms a magnetosphere, shielding the planet from the incoming solar wind particles. 
Magnetic reconnection and collisionless shocks are two phenomena that require a kinetic description to model the dissipation mechanisms arising from wave-particle interactions. 
The presence and location of such phenomena in a small magnetosphere depend on the solar wind velocity and temperature, on the dipolar magnetic field and on the IMF. 
As an example of these simulations, we carried out a three-dimensional simulation of a magnetosphere. The simulation set-up, parameters and main results are reported in \citet{Peng:etal:2015PCS}. 
In this simulation, the solar wind is sub-sonic as the solar wind velocity is lower than the magneto-sonic Mach number. As a result, no bow shock is forming but a turbulent area is present at the day side. 
This is clearly seen in the magnetic field lines, represented with grey tubes in the two panels of Figure~\ref{fig:magnetosphere}. 
In the simulation, the IMF points northward and magnetic reconnection is observed at high-latitude but not at the day-side magnetopause or magnetotail. 
A Kelvin-Helmholtz instability develops on the equatorial plane at the magnetopause flanks \citep{hasegawa2004transport}. 
This is visible from the electron density contour-plot on the XY plane, and enhanced energy conversion (left panel of Figure~\ref{fig:magnetosphere}). 

\subsection{Results: dipole}

\begin{figure}
\centering
\includegraphics[width=1\columnwidth]{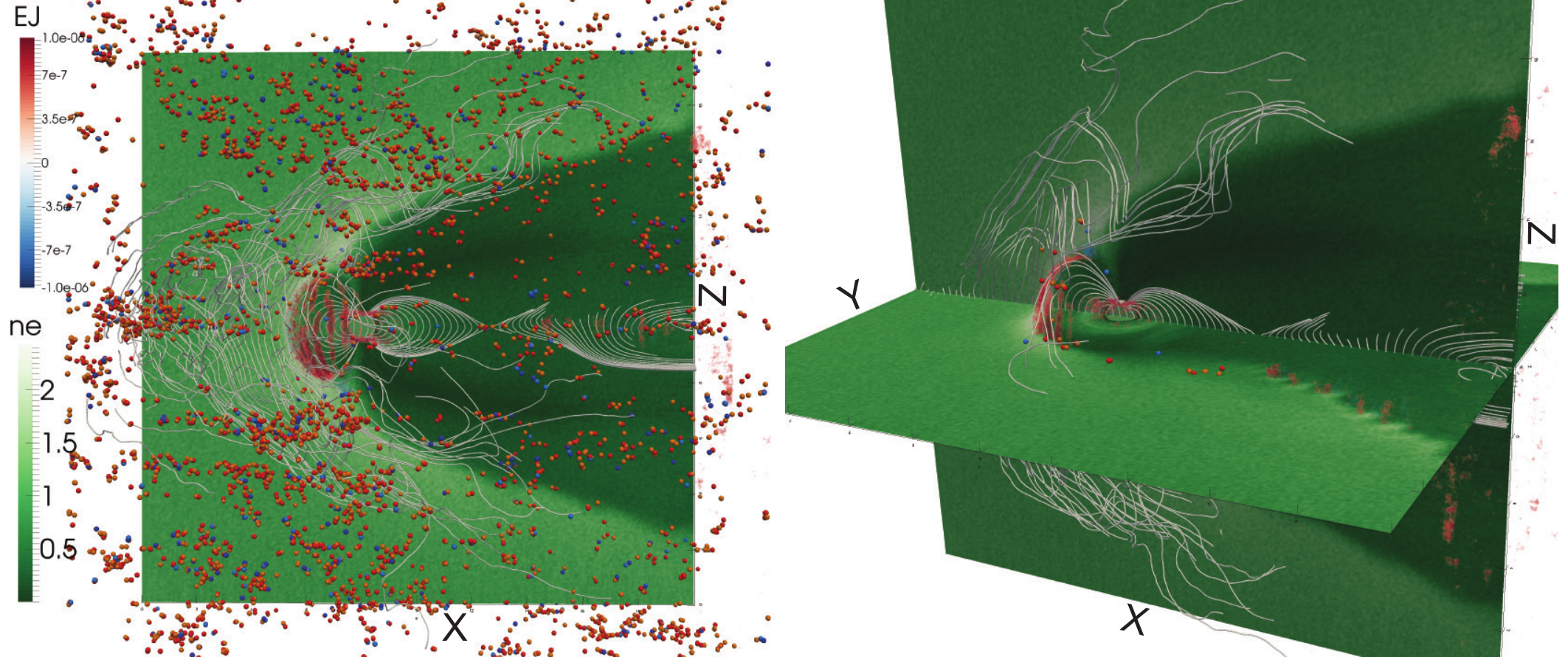}
\caption{Contour plot of the electron density on the XZ (left panel) and XZ-XY (right panel) planes with superimposed magnetic field lines (grey tubes) and null points (color spheres). 
The color indication of the nulls is conventional. 
Blue-red fog-like volume rendering of $\mathbf{E}\cdot\mathbf{j}$ indicates the regions of intense energy conversion.
}\label{fig:magnetosphere}
\end{figure}

More than eight thousand nulls have been detected in the last snapshot of this simulation (see Table~\ref{tab:allnulls} and Figure~\ref{fig:magnetosphere}a). 
The majority of nulls are located between the injection plane of the solar wind and the planet magnetosphere. 
Therefore, the null points are associated to the region with turbulent magnetic field arising form the interaction of the solar wind with the planet's magnetic dipole. 
No signatures of large-scale electromagnetic energy dissipation events are observed in this region.

However, close to the bow shock the regions of intense $\mathbf{E}\cdot\mathbf{j}$ are distributed in elongated stripes with alternating sign (Fig.~\ref{fig:magnetosphere}a).
Inverse energy transfer (negative $\mathbf{E}\cdot\mathbf{j}$) regions are weaker, and are barely noticeable.
In the right panel of Figure~\ref{fig:magnetosphere} the nulls outside a sphere centered on the planet have been clipped for clarity.
Few tens of magnetic nulls are visible in the magnetosheath and the magnetotail. 
Those on the dayside are close to the energy conversion regions.
Energy dissipation in the magnetotail is prominent in the turbulent structures formed by the Kelvin-Helmholtz instability.

Simulation of the dipolar planetary magnetosphere has revealed thousands of magnetic nulls, the dominant majority of which are of spiral type, well in accordance with the previously reported simulations.
Most of nulls form in the turbulence of the low guide field solar wind and demonstrate no major energy dissipation.
Only some spiral nulls in the magnetosheath or the bow shock are associated with the energy dissipation processes at the solar wind -- magnetosphere interface.

\section{Quadrupolar mini-magnetosphere}
\label{sec:divin}
The kinetic interaction of a non-dipolar magnetic field with the solar wind is discussed briefly in this section. 
An example of such topology is the paleomagnetosphere of the Earth, a configuration which possibly occurs during magnetic pole reversals \citep{McFadden2000}, a process estimated to take from 1,000 to 10,000 years each 0.1--1 million years.

Due to difficulties in reconstructing the field topology during the transition, it is yet a matter of debate of how intense the higher-order multipoles were at that time. 
It is hypothesized that the total dipole magnetic field energy is transferred to quadrupole or octupole moments \citep{Vogt2000,Zieger2004}. 
Therefore the quadrupole component can be strong enough to generate a fully-pledged magnetosphere, as well as to prevent the solar wind from directly impacting the atmosphere (at least in some parts of the globe). 
To make our simulations feasible, we scale down the quadrupole moment to study the mini-magnetosphere \citep{Lin_etal_1998} on the size of several ion inertial lengths. 
Although such object is hardly comparable to the real terrestrial magnetosphere, it is yet an important step in constructing the fully kinetic model with realistic scales.

Similar to a mini-magnetosphere formed by a dipolar source (Section~\ref{sec:bopeng}), the quadrupolar field creates cavity in the solar wind by deflecting solar wind particles. 
The resulting configuration is rather different \citep{Vogt2004} and contains several cusp regions, as well as several distinct current systems.
As this simulation has never been reported before, we introduce it in more details than the previously studied simulations.

\subsection{Quadrupolar initial condition}
The magnetic field $\textbf{B}$ can be written using the potential 
\begin{equation}
\Psi = \frac{1}{2} \frac{\textbf{x}^T \hat{Q} \textbf{x} } {x^5}, \;\;\; \textbf{B}=-\nabla \Psi,
\label{equation_1}
\end{equation}
where $\hat{Q}$ is the quadrupole tensor. By rotating the tensor to an appropriate basis with respect to which $\hat{Q}$ is diagonal [Vogt, 2004], one gets:
\begin{equation}
\hat{Q}=q
  \begin{bmatrix}
    -(1- \eta)/2&0&0 \cr
    0&-(1+ \eta)/2&0 \cr
    0&0&1 \cr
  \end{bmatrix}
\label{equation_2}
\end{equation}
Here, $q$ is the quadrupole strength, and $\eta \in [-1;1]$ is the ``shape'' parameter. 
In the given form (eq. \ref{equation_1}, \ref{equation_2}) the quadrupole axis is in the Z direction. 
Rotation around an axis can be implemented to tilt the source (see Table \ref{tab:magnetosphere} for runs 1--4 details).

The simulational domain size is $L_x \times L_y \times L_z = 18 \; d_i \times 9 \; d_i \times 9 \; d_i$, where the ion inertial length $d_i$ is computed for the upstream solar wind density. 
The density $n_0=1$ is set uniform in the domain at $t=0$. 
Solar wind is injected through one boundary, while all other boundaries are open.
The center of the planet and quadrupole source are located at $x=7\;d_i$, $y=4.5\;d_i$, $z=4.5\;d_i$; the incident solar wind velocity is $v_{\rm SW}/c=0.025$, and the thermal velocities of ions and electrons are, respectively, $0.0063$ and $0.045$. 
The sphere of radius $0.5$ simply absorbs particles, ionospheric effects are not included. 
The quadrupole source and solar wind parameters can be found in Table \ref{tab:magnetosphere}, other parameters are presented in Table~\ref{tab:allruns}.

\subsection{Results: quadrupole}

\begin{figure}
\includegraphics[width=1\textwidth]{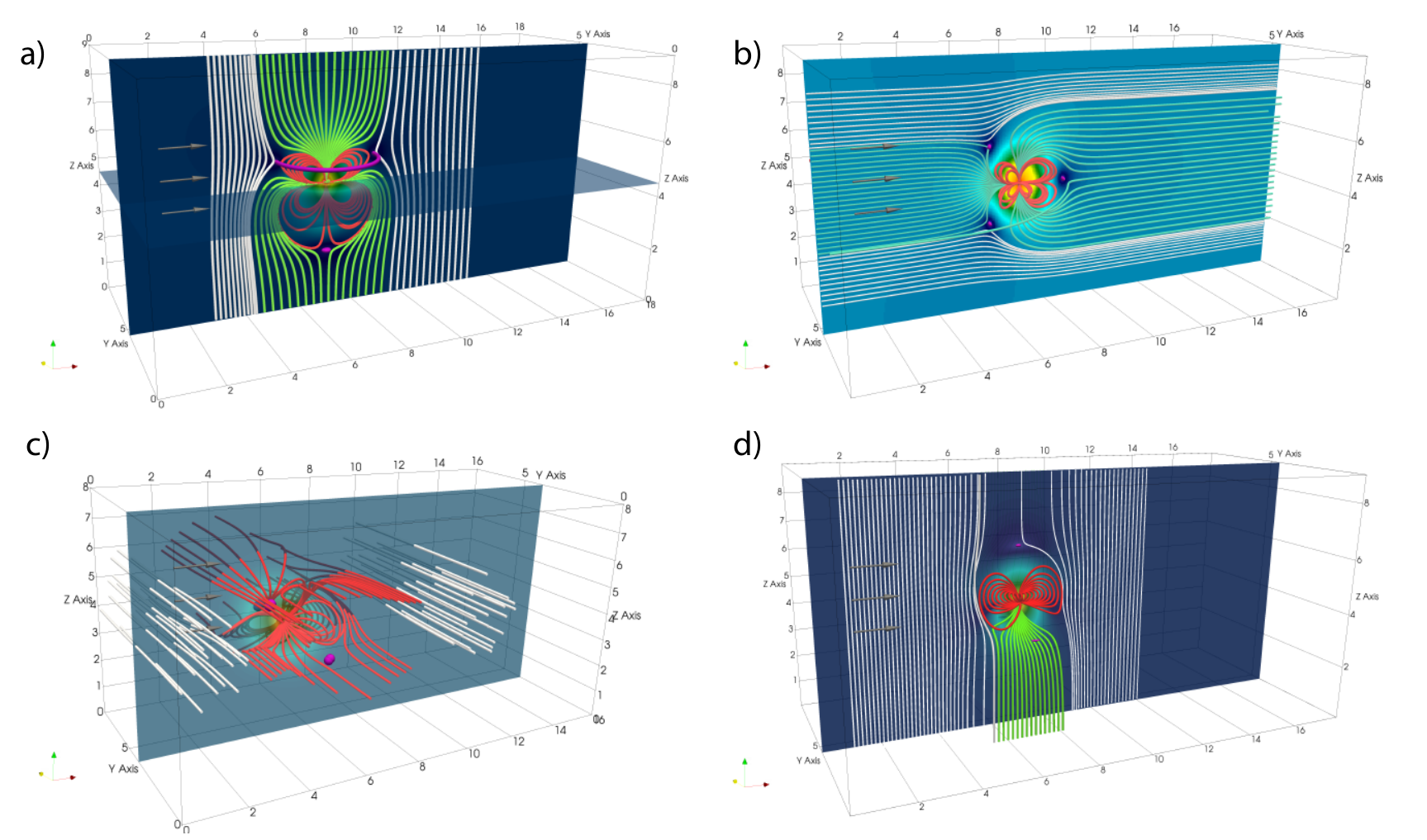}
\caption{Initial magnetic field configuration (a superposition of the quadrupole field with an external magnetic field) for runs 1--4 (a--d). 
Closed field lines are marked red. Open field lines are marked green (a, b, d). Solar wind field lines are white. Purple contours mark the regions of low magnetic field $B<1/20 B_{SW}$ enclosing the initial null points.}
\label{fig:divin1}
\end{figure}

Figure \ref{fig:divin1} displays the magnetic field configuration at $t=0$: a superposition of the quadrupole field (eq. \ref{equation_1}, \ref{equation_2}) and the solar wind magnetic field. 
To guide the eye, field lines are colored according to their connectivity: solar wind field lines are shown in white, field lines which start and end at the planet are red, semi-open field lines are green (Figure~\ref{fig:divin1}a, \ref{fig:divin1}b, \ref{fig:divin1}d). 
In the axisymmetric case ($\eta=0$, Runs 1--3) the quadrupole topology reminds that of a pair of stacked oppositely-directed dipoles, with cusp regions located at the poles and in the equatorial plane. 
The addition of an arbitrary uniform external field creates three null-points in Runs 2--4 \citep{Leubner:Zollner:1985}, which are visible in Figures~\ref{fig:divin1}b, \ref{fig:divin1}c, \ref{fig:divin1}d.
 
An important limiting case appears if $\mathbf{B}_{\rm SW}$ is parallel (or antiparallel) to the quadrupole axis. 
Such configuration has a single null in the `closed' half of the quadrupolar magnetosphere, accompanied by a line of nulls in the `open' half (Fig. \ref{fig:divin1}a, compare to Fig.~\ref{fig:LMA}). 
Therefore, in such a case the formed mini-magnetosphere shows features of both open and closed dipolar magnetospheres simultaneously.

\begin{figure}
\includegraphics[width=1\textwidth]{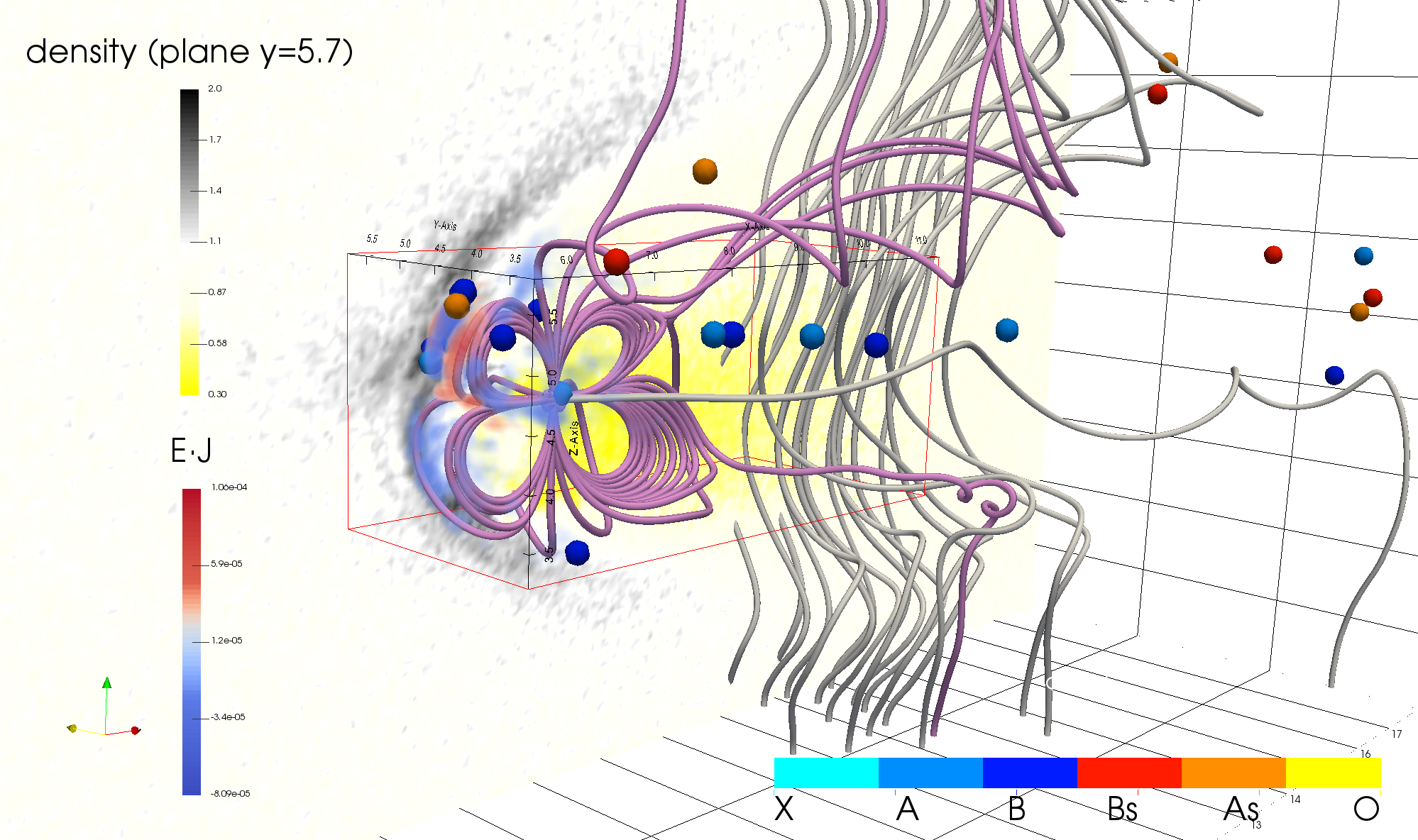}
\caption{Magnetic nulls and energy exchange at cycle 12000. Null coloring is identical to that of Figure~\ref{fig:lapenta1}. 
Field lines connected to the source are shown in magenta. Solar wind magnetic field lines are white. 
Plasma density distribution in the plane $Y=5.7 d_i$ is displayed with the black-yellow colormap. Volume rendering (fog-like) of $\textbf{E} \cdot \textbf{J}$ is displayed with the blue-red colormap.}
\label{fig:divin2}
\end{figure}

A halo of compressed plasma surrounds the mini-magnetosphere. The peak density ($n/n_{SW} \sim 2.5$ is reached at the dayside magnetopause (Fig. \ref{fig:divin2}, indicated in dark-gray). 
The solar wind flow drags magnetic field lines to the night side thus creating a tail with reduced plasma density.

A single A-type null point resides in the closed (lower) part of the quadrupole nearly at the same location as at $t=0$ (compare to Fig. \ref{fig:divin1}a). 
Energy exchange $\textbf{E} \cdot \textbf{J}<0$ is rather low and peaks at the subsolar point and not in the null. 
Such single null configuration is similar to the separator magnetopause reconnection occuring under northward IMF in a closed magnetosphere \citep{Dorelli2007}. 
In the open (upper) hemisphere, a ring of nulls disappears producing a collection of A- and B-type nulls, contrary to the solar wind -- dipole interaction, where most nulls in the magnetosphere are spiral (see Section \ref{sec:deca}, Fig.~\ref{fig:LMA}). 
Energy exchange is of both signs and is not focused exclusively in the nulls.

The interaction of the quadrupolar magnetosphere with the solar wind is similar to the dipole magnetosphere case (Section~\ref{sec:bopeng}).
In particular, the characteristic stripes of intense energy conversion are observed in the dayside magnetosphere.
However, more like in the LMA simulation (Section~\ref{sec:deca}) a substantial amount of radial nulls (that do not exhibit enhanced energy dissipation) are created here.

\section{Summary}

\begin{deluxetable}{lrrrrcl}
\tablecolumns{7}
\tablewidth{0pc}
\tablecaption{Summary of the nulls found in all simulations}
\tablehead{
  \colhead{Run} & \colhead{$A$} & \colhead{$B$}   & \colhead{$A_s$} & \colhead{$B_s$}    & $A_s+B_s$ (\%)  &  \colhead{Energy dissipation?}    
}
\startdata
Harris $B_g=0$   &   3  &  10  &  33  &   32  &  83   &  Flux ropes and current filaments.  \\
Harris $B_g=0.1$ &   1  &   0  &   5  &    6  &  92   &  Flux ropes and current filaments.  \\
Asymmetric       &136  & 141  & 439  &  428  &  76    &  Flux ropes and current filaments. \\   
Multiple nulls   &   5  &   5  &   54 &   58  &  92   &  Flux ropes (Z-pinches).  \\
Dipole           & 584  & 577  & 3836 & 3778  &  87   &  Bow shock, magnetopause, magnetotail.  \\
Quadrupole       &   9  &   9  &    9 &   12  &  54   &  Bow shock, magnetopause.  \\
LMA              &  10  &  11  &   2  &    0  &   9   &  No reconnection; radial null line.  \\
\enddata
\vspace{-0.8cm}
\tablecomments{Misdetections of nulls and their types is possible, therefore the topological ``parity'' is not conserved: sometimes positive nulls do not have their negative counterparts, and vice versa.}
\label{tab:allnulls}
\end{deluxetable}

We have analyzed seven different kinetic simulations of magnetized space plasmas: Harris current sheet with and without guide field, asymmetric reconnection, relaxing configuration with multiple nulls, lunar magnetic anomaly (LMA), dipolar and quadrupolar planetary mini-magnetospheres.
We have investigated magnetic nulls and their relation to the magnetic energy dissipation.
Our findings are summarized in Table~\ref{tab:allnulls}.

By the number of detected nulls the simulations fall into two categories: in the LMA and quadrupolar mini-magnetosphere the number of radial nulls overwhelms or compares to the number of spiral nulls.
In the LMA simulation a magnetic dipole interacts with the magnetic field of the solar wind, parallel to the axis of the dipole.
Under these conditions a line of radial nulls forms naturally on the interface.
As the size of the dipole is compared to ion diffusion scales, no indicators of magnetic reconnection or energy dissipation are observed.
The axis of the quadrupolar mini-magnetosphere studied here is parallel to the magnetic field of the solar wind. 
Such configuration has features of both open and closed dipolar magnetospheres.
In both cases the topology is governed rather by the boundary conditions, then by the plasma itself.

In all other cases the fraction of spiral nulls is over 75\%, well in agreement with the recent observational survey \citep{Eriksson:etal:2015}.
Our results are influenced by the artificial symmetries in most of the initial configurations. 
For instance, extended null surfaces in the current sheets with no guide field are topologically unstable.
In reality such current sheet should break into many separated nulls which topology should depend on the surrounding currents.
On the other hand, our finding is a reflection of the common physical processes that take place in various magnetized space plasmas.
Topological bifurcations that lead to the creation of magnetic nulls in turbulent plasmas have similar nature, be it a pristine (weakly magnetized) solar wind, magnetosheath, or a reconnection outflow in the magnetotail.

Introduction of a small guide field into the magnetic reconnection simulation dramatically decreases the number of nulls detected in the reconnection outflow.
Perhaps this finding could be used as an additional indicator of the presence or absence of a guide field in the observations.

Often, magnetic nulls are poor proxies for the energetic events.
However, short-living nulls accompanied by enhanced currents indicate magnetic reconnection or energy release events.
Such nulls are created by topological bifurcations caused by interacting magnetic flux ropes.

Magnetic flux ropes enclose current filaments, resembling magnetic islands or pinches, and often have spiral nulls at their axes.
These nulls are connected by twisted magnetic field lines, and, alternating in sign, form spiral null lines.
Presence of a spiral null line hints that energy dissipation is happening nearby, within a distance of the ion inertial length.
The dissipation is driven by the fundamental instabilities occuring in the current filaments themselves and during their interaction.

We foresee two important implications of this work:
\begin{enumerate}
\item Observers should neither completely disregard nulls, nor focus exclusively on magnetic topology. It is important to invoke information about other plasma properties such as flows and particle density. It would be very interesting to perform a similar study in solar plasmas (where no initial symmetry is present): in the lower atmosphere (photosphere, chromosphere) where vector magnetograms are available, and in the extrapolated coronal fields.
\item Our results dictate a rather unusual concept of energy dissipation in turbulent space plasmas with a small guide field, driven by current filaments and magnetic flux ropes. More insight should be given to the instabilities that govern energy exchange in such current filaments and during their interaction. Is it possible to extend this concept to the configurations with a strong guide field and to the solar plasma? This question is, in our opinion, of major importance for the future of numerical simulations and observations of space and solar plasmas.
\end{enumerate}

\acknowledgments

The work is supported by the Onderzoekfonds KU Leuven (Research Fund KU Leuven).
V.O. is supported by the Air Force Office of Scientific Research, Air Force Materiel Command, USAF under Award No. FA9550-14-1-0375.
J.D. is supported by NASA's Solar System Exploration Research Virtual Institutes's (SSERVI) Institute for Modeling Plasmas, Atmospheres, and Cosmic Dust (IMPACT).
M.E.I. is supported by the FWO (Fonds Wetenschappelijk Onderzoek Vlaanderen) postdoctoral fellowship (12O5215N).
G.L. acknowledges support from the NASA MMS Grant No. NNX08AO84G.
This research has received funding from the European Commission's FP7 Program with the grant agreement DEEP-ER (project ICT-610476, http://www.deep-er.eu/).
The simulations were conducted on the computational resources provided by the PRACE Tier-0 projects 2011050747 (Curie) and 2013091928 (SuperMUC).
Great part of this work was done during the Nordita program on Magnetic Reconnection in Plasmas 2015. 
Authors are thankful to Mikhail Sitnov for useful discussions.

\appendix


\begin{deluxetable}{lcclrlrlccc}
\rotate
\tablewidth{0pc}
\tablecaption{All simulation runs}
\tablehead{
  \colhead{Run} & \colhead{Dimensions ($d_i$)} & \colhead{Number of cells}   & \colhead{$\frac{dx}{d_i}$} & \colhead{$\frac{m_i}{m_e}$}    & \colhead{$dt\omega_{pi}$}  &  \colhead{$T\omega_{pi}$}   & \colhead{$\frac{\upsilon_{th,i}}{c}$}    & \colhead{$\frac{\upsilon_{th,e}}{c}$} & \colhead{$N_s$} & \colhead{$N_p$}
}
\startdata
Harris $B_g=0$   &  $40\times 15\times 10$           & $512\times 192\times 128$        & 0.078        &  256   &  0.125   &  2000  &  0.003    &  0.045   &  4  &  125  \\
Harris $B_g=0.1$ &  $40\times 15\times 10$           & $512\times 192\times 128$        & 0.078        &  256   &  0.125   &  2000  &  0.003    &  0.045   &  4  &  125  \\
Asymmetric       &  $20.16\times 20.16\times  6.09$  & $516\times 516\times 156$        & 0.039        &  256   &  0.5     &  4500  &  0.00046  &  0.005   &  2  &   25  \\
LMA              &  $0.625\times 1.25\times 1.25$    & $320\times 640\times 640$        & 0.002        &  256   &  0.019   &   188  &  0.003    &  0.045   &  2  &   64  \\
Dipole           &  $20\times 20\times 20$           & $384\times 384\times 384$        & 0.052        &   64   &  0.15    &  2250  &  0.011    &  0.078   &  2  &   64  \\
Quadrupole       &  $18\times  9\times  9$           & $512\times 192\times 128$        & 0.078        &  256   &  0.075   &   900  &  0.006    &  0.045   &  2  &  100  \\
Multiple nulls   &  $20\times 20\times 20$           & $400\times 400\times 400$        & 0.05         &   25   &  0.15    &  2250  &  0.009    &  0.020   &  2  &   64  \\
\enddata
\vspace{-0.8cm}
\tablecomments{$N_p$ is the number of particles/specie/cell. $N_s$ is the number of species.}
\label{tab:allruns}
\end{deluxetable}

\begin{deluxetable}{rrrrcr}
\tablecolumns{6}
\tablewidth{0pc}
\tablecaption{Mini-magnetosphere simulations}
\tablehead{
  \colhead{Run} & \colhead{$q$} & \colhead{$\eta$}   & \colhead{$B_{SW}=[B_x,\;B_y,\;B_z]$} & \colhead{Rotation (axis)}  & \colhead{SW speed $\upsilon_{SW}/c$}
}
\startdata
\cutinhead{LMA}
Run 1 &$0.0005$ &   -      &  $[0.0,\;0.0016,\;0.0]$    & -  &  0.017  \\
\cutinhead{Dipole}
Run 1 & -       &   -      &  $[0.0001,\;0.0,\;0.0]$    & -  &  0.02  \\
\cutinhead{Quadrupole}
Run 1 & $0.05$  & $0.0$    &  $[0.0,\;0.0,\;0.002]$     & -   &  0.025 \\
Run 2 & $0.05$  & $0.0$    &  $[0.005,\;0.0,\;0.0]$     & -   &  0.025 \\
Run 3 & $0.05$  & $0.0$    &  $[0.0,\;0.003,\;0.0]$     & $45\degree$ (y axis) &  0.025 \\  
Run 4 & $0.05$  & $1.0$    &  $[0.0,\; 0.0, \; -0.005]$ & -   &  0.025 \\  
\enddata
\label{tab:magnetosphere} 
\end{deluxetable}

\bibliographystyle{apj}
\bibliography{main,cazzola_innocenti,deca,divin,bopeng}

\end{document}